# Geodesic equations and their numerical solutions in geodetic and Cartesian coordinates on an oblate spheroid

Georgios Panou, Romylos Korakitis

Department of Surveying Engineering, National Technical University of Athens, Zografou Campus, 15780 Athens, Greece (geopanou@survey.ntua.gr)

**Abstract** The direct geodesic problem on an oblate spheroid is described as an initial value problem and is solved numerically in geodetic and Cartesian coordinates. The geodesic equations are formulated by means of the theory of differential geometry. The initial value problem under consideration is reduced to a system of first-order ordinary differential equations, which is solved using a numerical method. The solution provides the coordinates and the azimuths at any point along the geodesic. The Clairaut constant is not assumed known but it is computed, allowing to check the precision of the method. An extended data set of geodesics is used, in order to evaluate the performance of the method in each coordinate system. The results for the direct geodesic problem are validated by comparison to Karney's method. We conclude that a complete, stable, precise, accurate and fast solution of the problem in Cartesian coordinates is accomplished.

## 1 Introduction

In geodesy, there are two traditional problems concerning geodesics on an oblate spheroid (ellipsoid of revolution): (i) the direct problem: given a point $P_0$ on an oblate spheroid, together with the azimuth $\alpha_0$ and the geodesic distance $s_{01}$ to a point $P_1$, determine the point $P_1$ and the azimuth $\alpha_1$ at this point, and (ii) the inverse problem: given two points $P_0$ and $P_1$ on an oblate spheroid, determine the geodesic distance $s_{01}$ between them and the azimuths $\alpha_0$, $\alpha_1$ at the end points.

These problems have a very long history and several different methods of solving them have been proposed by many researchers, as they are reported in a comprehensive list by Karney (2016) in GeographicLib. Also, some of the existing methods have been presented by Rapp (1993) and Deakin and Hunter (2010).

The methods of solving the above problems can be divided into two general categories: (i) using an auxiliary sphere, e.g., Bessel (1826), Rainsford (1955), Robbins (1962), Sodano (1965), Saito (1970), Vincenty (1975), Saito (1979), Bowring (1983), Karney (2013) and (ii) without using an auxiliary sphere, e.g., Kivioja (1971), Holmstrom (1976), Jank and Kivioja (1980), Thomas and Featherstone (2005), Panou (2013), Panou et al. (2013), Tseng (2014). The methods which use an auxiliary sphere are based on the classical work of Bessel (1826) and its modifications. On the other hand, the methods without using an auxiliary sphere are attacking the problems directly on an oblate spheroid. They are conceptually simpler and can be generalized in the case of a triaxial ellipsoid, as has been already presented by Holmstrom (1976) and Panou (2013).

The solution of the geodesic problems, with one of the above two methods, includes evaluating elliptic integrals or solving differential equations using: (i) approximate analytical methods, e.g., Vincenty (1975), Holmstrom (1976), Pittman (1986), Mai (2010), Karney (2013) or (ii) numerical methods, e.g., Saito (1979), Rollins (2010), Sjöberg (2012), Sjöberg and Shirazian (2012), Panou et al. (2013). The approximate analytical methods are usually based on the fact that an oblate spheroid deviates slightly from a sphere, so these methods essentially involve a truncated series expansion. On the other hand, numerical methods can be used for ellipsoids of arbitrary flattening. In addition, they do not require a change in the theoretical background with a modification of the computational environment. However, they may suffer from computational errors, which are reduced with the improvements in modern computational systems. Furthermore, since the numerical methods perform computations at many points along the geodesic, they can be used as a convenient and efficient approach to trace the geodesic. If tracing is not needed, an analytical method may be sufficient to give the results of the geodesic problems.

From the vast literature on geodesic problems, one notices that the evaluation of the performance of the geodesic algorithms is based on the aspects of stability, accuracy and computational speed. Of course, the execution time depends strongly on the programming environment and the computing platform used. Today, with the broad availability of high speed computers, the execution time does not longer play an essential role. With regard to stability, the algorithms should be stable in the domain of use, i.e. without limitations to the input data of the problem. Finally, they should provide results with high accuracy, depending on the demands of the application.

In this work, the geodesic equations (independent variable $s$) are numerically solved directly on an oblate spheroid using two coordinate systems: geodetic and Cartesian. The presented method can be generalized in the case of a triaxial ellipsoid and can be used for arbitrary flattening. However, in order to evaluate the method, we limit the numerical applications to the case of the WGS84 oblate spheroid.

In general, there are several numerical methods for solving an initial value problem (see Hildebrand 1974). In this study, we will use a relatively simple and commonly applied fourth-order Runge-Kutta method (see Butcher 1987), which has been applied successfully to the geodesic boundary value problem (Panou 2013, Panou et al. 2013). In addition, we examine the performance (stability, precision, execution time) of the method in both coordinate systems.

Kivioja (1971) has solved the geodesic initial value problem by numerical integration of a system of two differential equations. Thomas and Featherstone (2005) improved Kivioja's method by altering the system of the two differential equations along the geodesic, in order to avoid the singularity when the geodesic passes through the vertices. Alternatively, in this study we propose the numerical integration of a system of four differential equations of a geodesic in geodetic coordinates, which is free from this singularity. Although there are more equations, from the solution we can determine the Clairaut constant, at any point along the geodesic, and thus we are able to check the precision of the numerical integration. We should mention that other methods for the solution of the direct geodesic problem, such as Rollins (2010) and Sjöberg and Shirazian (2012), despite the fact that use the Clairaut constant as an a priori known quantity in the equations, they employ iterative techniques, so they demand a lot of computational effort.

Inevitably, in curvilinear coordinates there are two poles (singularities) and hence all of the above methods can be ill-behaved. This does not happen in Cartesian coordinates and the algorithms based on them are insensitive to singularities, such as $\tan(\pi/2)$. Also, the Cartesian coordinate system can be easily related to other curvilinear systems, many formulas in this system are simpler, without numerical difficulties and computations do not demand the use of trigonometric functions, which can make computer processing slow, as pointed out by Felski (2011).

Using the calculus of variations, the geodesic equations in Cartesian coordinates were derived on a sphere by Fox (1987) and on a triaxial ellipsoid by Holmstrom (1976). Although the approximate analytical solution given by Holmstrom (1976) can be applied in the degenerate case of an oblate spheroid, it is of low precision, since the precision was not his primary consideration.

Part of the numerical solution of the geodesic initial value problem is the solution of the direct geodesic problem, where the position and the azimuth are determined only at the end point of the geodesic. Although today, with Global Navigation Satellite System (GNSS) technologies, the inverse geodesic problem is more realistic than the direct geodesic problem, the proposed algorithm can be used iteratively for the solution of the inverse problem, as has already been suggested by Jank and Kivioja (1980) and Vermeer (2015). Finally, the proposed method is an independent method which can be used to validate Karney's method (Karney 2013) for the direct geodesic problem.

## 2 Geodesics in geodetic coordinates

The geodesic initial value problem, expressed in geodetic coordinates on an oblate spheroid, consists of determining a geodesic, parametrized by its arc length $s$, $\varphi = \varphi(s)$, $\lambda = \lambda(s)$, with azimuths $\alpha = \alpha(s)$ along it, which passes through a given point

$P_0(\varphi(0), \lambda(0))$ in a known direction (given azimuth $\alpha_0 = \alpha(0)$) and has a certain length $s_{01}$.

## 2.1 Geodesic equations

We consider an oblate spheroid which, in geodetic coordinates ($\varphi$, $\lambda$), is described parametrically by

$$x = N \cos\varphi \cos\lambda \tag{1a}$$
$$y = N \cos\varphi \sin\lambda \tag{1b}$$
$$z = N(1 - e^2)\sin\varphi \tag{1c}$$

where $\varphi$ ($-\pi/2 \leq \varphi \leq +\pi/2$) is the geodetic latitude, $\lambda$ ($-\pi < \lambda \leq +\pi$) is the geodetic longitude and

$$N = \frac{a}{(1 - e^2 \sin^2\varphi)^{1/2}} \tag{2}$$

is the radius of curvature in the prime vertical normal section, with $a$ the major semiaxis and $e$ the first eccentricity. Also, it holds that $e^2 = f(2 - f)$, where $f$ is the flattening. In this parametrization, the elements of the first fundamental form are (Vermeer 2015)

$$E = \frac{a^2(1 - e^2)^2}{(1 - e^2 \sin^2\varphi)^3} \tag{3a}$$
$$F = 0 \tag{3b}$$
$$G = \frac{a^2 \cos^2\varphi}{1 - e^2 \sin^2\varphi} \tag{3c}$$

In Eq. (3b), $F = 0$ indicates that the $\varphi$-curves (parallels) and $\lambda$-curves (meridians) are orthogonal. Also, $E \neq 0$ for all $\varphi$ and $G = 0$ when $\varphi = \pm\pi/2$ (at the poles) (Panou et al. 2013). From Eqs. (3), we obtain the derivatives

$$E_\varphi = \frac{3e^2 a^2 (1-e^2)^2 \sin(2\varphi)}{(1-e^2 \sin^2 \varphi)^4} \tag{4a}$$

$$F_\varphi = 0 \tag{4b}$$

$$G_\varphi = -\frac{a^2(1-e^2)\sin(2\varphi)}{(1-e^2\sin^2\varphi)^2} \tag{4c}$$

$$E_\lambda = F_\lambda = G_\lambda = 0 \tag{5}$$

Thus, the Christoffel symbols $\Gamma^i_{jk}$ ($i, j, k = 1, 2$) become (Struik 1961, p. 107)

$$\Gamma^1_{11} = \frac{E_\varphi}{2E} = \frac{3e^2 \sin(2\varphi)}{2(1-e^2\sin^2\varphi)} \tag{6a}$$

$$\Gamma^1_{22} = -\frac{G_\varphi}{2E} = \frac{(1-e^2\sin^2\varphi)\sin(2\varphi)}{2(1-e^2)} \tag{6b}$$

$$\Gamma^2_{12} = \frac{G_\varphi}{2G} = -\frac{(1-e^2)\tan\varphi}{1-e^2\sin^2\varphi} \tag{6c}$$

$$\Gamma^1_{12} = \Gamma^2_{11} = \Gamma^2_{22} = 0 \tag{6d}$$

Therefore, the geodesic equations, expressed in geodetic coordinates on an oblate spheroid, are given by (Struik 1961, p. 132)

$$\frac{d^2\varphi}{ds^2} + \Gamma^1_{11}\left(\frac{d\varphi}{ds}\right)^2 + \Gamma^1_{22}\left(\frac{d\lambda}{ds}\right)^2 = 0 \tag{7a}$$

$$\frac{d^2\lambda}{ds^2} + 2\Gamma^2_{12}\frac{d\varphi}{ds}\frac{d\lambda}{ds} = 0 \tag{7b}$$

The initial conditions associated with these equations are

$$\varphi_0 = \varphi(0), \quad \left.\frac{d\varphi}{ds}\right|_0 = \frac{d\varphi}{ds}(0) \tag{8a}$$

$$\lambda_0 = \lambda(0), \quad \left.\frac{d\lambda}{ds}\right|_0 = \frac{d\lambda}{ds}(0) \tag{8b}$$

where the values of the derivatives at point $P_0(\varphi_0, \lambda_0)$ are given below. Hence, the direct geodesic problem is described as an initial value problem in geodetic coordinates on an oblate spheroid by Eqs. (7) and Eqs. (8).

**2.2 Numerical solution**

In order to solve the geodesic initial value problem by a numerical method, the system of two non-linear second-order ordinary differential equations (Eqs. (7)) is written equivalently as a system of four first-order differential equations:

$$\frac{d}{ds}(\varphi) = \frac{d\varphi}{ds} \tag{9a}$$

$$\frac{d}{ds}\left(\frac{d\varphi}{ds}\right) = -\Gamma^1_{11}\left(\frac{d\varphi}{ds}\right)^2 - \Gamma^1_{22}\left(\frac{d\lambda}{ds}\right)^2 \tag{9b}$$

$$\frac{d}{ds}(\lambda) = \frac{d\lambda}{ds} \tag{9c}$$

$$\frac{d}{ds}\left(\frac{d\lambda}{ds}\right) = -2\Gamma^2_{12}\frac{d\varphi}{ds}\frac{d\lambda}{ds} \tag{9d}$$

This system can be integrated on the interval [0, *s*] using a numerical method, such as Runge-Kutta (see Hildebrand 1974, Butcher 1987). The step size δ*s* is given by δ*s* = *s*/*n*, where *n* is the number of steps. As a rule, a greater number of steps leads to a greater precision but also to a greater execution time and vice versa. However, the effects of the number of steps and the performance of the method (stability, precision, execution time) are examined in detail in Section 4.

For the variables φ and λ, the initial conditions are $\varphi_0$ and $\lambda_0$, respectively. For the required derivatives, we recall the well-known relations of a differential element on an oblate spheroid, for any curve of length *ds* (Vermeer 2015)

$$\frac{d\varphi}{ds} = \frac{\cos\alpha}{M} \tag{10a}$$

$$\frac{d\lambda}{ds} = \frac{\sin\alpha}{N\cos\varphi} \qquad (10b)$$

where

$$M = \frac{a(1-e^2)}{(1-e^2\sin^2\varphi)^{3/2}} \qquad (11)$$

is the radius of curvature in the meridian normal section. Hence, the required values of the derivatives at point $P_0(\varphi_0, \lambda_0)$ are

$$\left.\frac{d\varphi}{ds}\right|_0 = \frac{\cos\alpha_0}{M(\varphi_0)} \qquad (12a)$$

$$\left.\frac{d\lambda}{ds}\right|_0 = \frac{\sin\alpha_0}{N(\varphi_0)\cos\varphi_0} \qquad (12b)$$

**2.3 Azimuths and Clairaut's constant**

Using Eqs. (10), the azimuth α at any point along the geodesic is computed by

$$\alpha = \arctan\left(\frac{V}{U}\right) = \text{arccot}\left(\frac{U}{V}\right) \qquad (13)$$

where

$$U = M\frac{d\varphi}{ds} \qquad (14a)$$

$$V = N\cos\varphi\frac{d\lambda}{ds} \qquad (14b)$$

Note that Eq. (13) involves the variables φ, $d\varphi/ds$ and $d\lambda/ds$, which are obtained by the numerical integration.

The integration of Eq. (7b) yields

$$\frac{d\lambda}{ds}G = C \tag{15}$$

where $C$ is an arbitrary constant. We note that Eq. (15) involves only the variables $\varphi$ and $d\lambda/ds$. Substituting Eq. (10b) into Eq. (15) and using Eqs. (2) and (3c), we obtain

$$N\cos\varphi\sin\alpha = C \tag{16}$$

which is the well-known Clairaut's equation in geodetic coordinates. Hence, Eq. (15) and Eq. (16) are mathematically equivalent. Also, we can estimate, at any value of the independent variable $s$, the difference $\delta C = C - C_0$ between the computed value $C$ and the known value $C_0$ at point $P_0$, from the given $\varphi_0$ and $\alpha_0$, by means of Clairaut's equation (Eq. (16)). In this way, we can check the precision of the numerical integration, since the difference $\delta C$ should be zero meters at any point along the geodesic.

## 3 Geodesics in Cartesian coordinates

In a similar manner, the geodesic initial value problem, expressed in Cartesian coordinates on an oblate spheroid, consists of determining a geodesic, parametrized by its arc length $s$, $x = x(s)$, $y = y(s)$, $z = z(s)$, with azimuths $\alpha = \alpha(s)$ along it, which passes through a given point $P_0(x(0), y(0), z(0))$ in a known direction (given azimuth $\alpha_0 = \alpha(0)$) and has a certain length $s_{01}$.

### 3.1 Geodesic equations

We consider an oblate spheroid which is described in Cartesian coordinates $(x, y, z)$ by

$$S(x,y,z) \doteq x^2 + y^2 + \frac{z^2}{1-e^2} - a^2 = 0 \tag{17}$$

It is well-known, from the theory of differential geometry, that the principal normal to the geodesic must coincide with the normal to the oblate spheroid (Struik 1961, Deakin and Hunter 2010), i.e.

$$\frac{d^2x/ds^2}{\partial S/\partial x} = \frac{d^2y/ds^2}{\partial S/\partial y} = \frac{d^2z/ds^2}{\partial S/\partial z} = -m \tag{18}$$

From these equations, together with Eq. (17), it is possible to determine $x(s)$, $y(s)$, $z(s)$ and $m(s)$. Using Eq. (17), Eq. (18) become

$$\frac{1}{x}\frac{d^2x}{ds^2} = \frac{1}{y}\frac{d^2y}{ds^2} = \frac{1-e^2}{z}\frac{d^2z}{ds^2} = -2m \tag{19}$$

Differentiating Eq. (17), we have

$$x\frac{dx}{ds} + y\frac{dy}{ds} + \frac{z}{1-e^2}\frac{dz}{ds} = 0 \tag{20}$$

and a further differentiation yields

$$x\frac{d^2x}{ds^2} + y\frac{d^2y}{ds^2} + \frac{z}{1-e^2}\frac{d^2z}{ds^2} = -\left[\left(\frac{dx}{ds}\right)^2 + \left(\frac{dy}{ds}\right)^2 + \frac{1}{1-e^2}\left(\frac{dz}{ds}\right)^2\right] \tag{21}$$

Hence, from Eq. (19) and Eq. (21), we obtain

$$m = \frac{h}{2H} \tag{22}$$

where

$$H = x^2 + y^2 + \frac{z^2}{(1-e^2)^2} \tag{23a}$$

$$h = \left(\frac{dx}{ds}\right)^2 + \left(\frac{dy}{ds}\right)^2 + \frac{1}{1-e^2}\left(\frac{dz}{ds}\right)^2 \tag{23b}$$

Substituting Eq. (22) into Eq. (19), we obtain the geodesic equations in Cartesian coordinates on an oblate spheroid

$$\frac{d^2x}{ds^2} + \frac{h}{H}x = 0 \tag{24a}$$

$$\frac{d^2y}{ds^2} + \frac{h}{H}y = 0 \tag{24b}$$

$$\frac{d^2z}{ds^2} + \frac{h}{H}\cdot\frac{z}{1-e^2} = 0 \tag{24c}$$

which are subject to the initial conditions

$$x_0 = x(0), \quad \left.\frac{dx}{ds}\right|_0 = \frac{dx}{ds}(0) \tag{25a}$$

$$y_0 = y(0), \quad \left.\frac{dy}{ds}\right|_0 = \frac{dy}{ds}(0) \tag{25b}$$

$$z_0 = z(0), \quad \left.\frac{dz}{ds}\right|_0 = \frac{dz}{ds}(0) \tag{25c}$$

where the values of the derivatives at point $P_0(x_0, y_0, z_0)$ are given below. Hence, the direct geodesic problem is described as an initial value problem in Cartesian coordinates on an oblate spheroid by Eqs. (24) and Eqs. (25).

### 3.2 Numerical solution

In order to solve the above problem, the system of three non-linear second-order ordinary differential equations (Eqs. (24)) is rewritten as a system of six first-order differential equations:

$$\frac{d}{ds}(x) = \frac{dx}{ds} \tag{26a}$$

$$\frac{d}{ds}\left(\frac{dx}{ds}\right) = -\frac{h}{H}x \tag{26b}$$

$$\frac{d}{ds}(y) = \frac{dy}{ds} \tag{26c}$$

$$\frac{d}{ds}\left(\frac{dy}{ds}\right) = -\frac{h}{H}y \tag{26d}$$

$$\frac{d}{ds}(z) = \frac{dz}{ds} \tag{26e}$$

$$\frac{d}{ds}\left(\frac{dz}{ds}\right) = -\frac{h}{H}\cdot\frac{z}{1-e^2} \tag{26f}$$

This system can be integrated on the interval [0, s] by a numerical method. Again, the step size $\delta s$ is given by $\delta s = s/n$, where $n$ is the number of steps. For the variables $x$, $y$ and $z$, the initial conditions are $x_0$, $y_0$ and $z_0$, respectively. To obtain the required derivatives, we proceed to describe the unit vectors to a geodesic through a point $P(x, y, z)$ on an oblate spheroid (see Fig. 1).

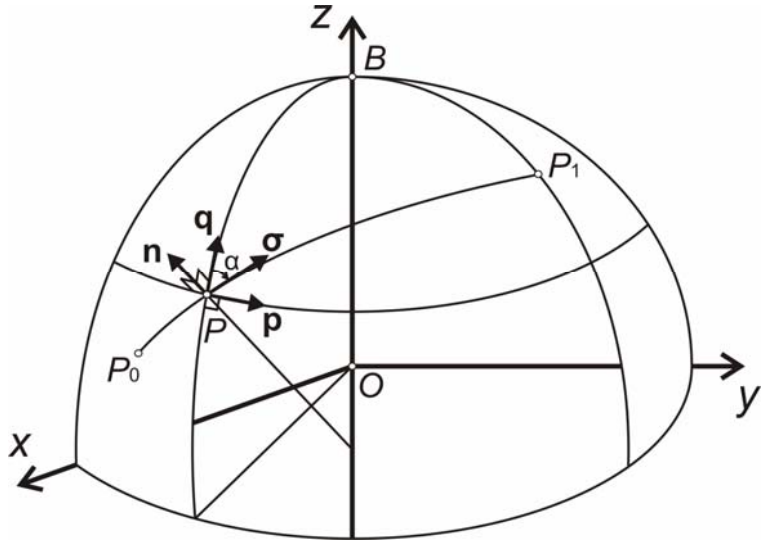

Figure 1. Unit vectors to a geodesic through a point $P$ on an oblate spheroid

Let $\boldsymbol{\sigma}$ be a unit vector tangent to an arbitrary geodesic through $P$. Then, we can express $\boldsymbol{\sigma}$ as (Fig. 1):

$$\boldsymbol{\sigma} = \left(\frac{dx}{ds}, \frac{dy}{ds}, \frac{dz}{ds}\right) = \mathbf{p}\sin\alpha + \mathbf{q}\cos\alpha \tag{27}$$

The unit vector normal to an oblate spheroid (using the gradient operator and Eqs. (17), (23a)) can be expressed as (Fig. 1):

$$\mathbf{n} = (n_1, n_2, n_3) = \left( \frac{x}{H^{1/2}}, \frac{y}{H^{1/2}}, \frac{z}{(1-e^2)H^{1/2}} \right) \tag{28}$$

Furthermore, considering the plane of the meridian of the point *P*, which passes also through the pole *B* and the centre of the oblate spheroid *O* (Felski 2011), we obtain the unit vector **p** (Fig. 1):

$$\mathbf{p} = (p_1, p_2, p_3) = \left( \frac{-y}{(x^2+y^2)^{1/2}}, \frac{x}{(x^2+y^2)^{1/2}}, 0 \right) \tag{29}$$

This vector has singularities at the poles, where we can simply set $\mathbf{p} = (0,1,0)$. Otherwise, this vector can be expressed in terms of geodetic longitude λ with the help of Eqs. (1):

$$\mathbf{p} = (p_1, p_2, p_3) = (-\sin\lambda, \cos\lambda, 0) \tag{30}$$

At the poles, we can now set λ = α, i.e. $\mathbf{p} = (-\sin\alpha, \cos\alpha, 0)$.

The unit vector **q**, tangent to the meridian, can now be determined as the cross product of unit vectors **n** and **p** (Fig. 1):

$$\mathbf{q} = \mathbf{n} \times \mathbf{p} = (q_1, q_2, q_3) = (-n_3 p_2, n_3 p_1, n_1 p_2 - n_2 p_1) \tag{31}$$

Finally, substituting the vectors **p** and **q** into Eq. (27), we obtain the required values of the derivatives at point $P_0(x_0, y_0, z_0)$

$$\left. \frac{dx}{ds} \right|_0 = p_1(0)\sin\alpha_0 + q_1(0)\cos\alpha_0 \tag{32a}$$

$$\left.\frac{dy}{ds}\right|_0 = p_2(0)\sin\alpha_0 + q_2(0)\cos\alpha_0 \tag{32b}$$

$$\left.\frac{dz}{ds}\right|_0 = p_3(0)\sin\alpha_0 + q_3(0)\cos\alpha_0 \tag{32c}$$

### 3.3 Azimuths and Clairaut's constant

Taking the scalar product of Eq. (27) successively with **p** and **q** and dividing the resulting equations, yields

$$\alpha = \arctan\left(\frac{R}{Q}\right) = \operatorname{arccot}\left(\frac{Q}{R}\right) \tag{33}$$

where

$$Q = \mathbf{q} \cdot \boldsymbol{\sigma} = q_1 \frac{dx}{ds} + q_2 \frac{dy}{ds} + q_3 \frac{dz}{ds} \tag{34a}$$

$$R = \mathbf{p} \cdot \boldsymbol{\sigma} = p_1 \frac{dx}{ds} + p_2 \frac{dy}{ds} + p_3 \frac{dz}{ds} \tag{34b}$$

Note that Eq. (33) involves all the variables $x$, $dx/ds$, $y$, $dy/ds$, $z$ and $dz/ds$, which are obtained by the numerical integration.

From the first equation of Eq. (19), a second-order differential equation, we obtain a first integral

$$x\frac{dy}{ds} - y\frac{dx}{ds} = C \tag{35}$$

where $C$ is an arbitrary constant. We note that Eq. (35) involves only the variables $x$, $dx/ds$, $y$ and $dy/ds$. Now, from the scalar product of Eq. (27) with **p** and using Eqs. (29), (35), we obtain

$$(x^2 + y^2)^{1/2} \sin\alpha = C \tag{36}$$

which is the well-known Clairaut's equation in Cartesian coordinates. Hence, Eq. (35) and Eq. (36) are mathematically equivalent. Also, in a manner similar to the geodetic coordinates, we can estimate, at any value of *s*, the difference $\delta C = C - C_0$ between the computed value *C* and the known value $C_0$ at point $P_0$, from the given $x_0$, $y_0$ and $\alpha_0$, by means of Clairaut's equation (Eq. (36)). Furthermore, because the numerical integration is performed in space, we can compute, at any value of *s*, the function *S*, given by Eq. (17). Therefore, we can double check the precision of the numerical integration, since the difference $\delta C$ should be zero meters and the function *S* should be zero at any point along the geodesic on an oblate spheroid.

**4 Numerical tests and comparisons**

**4.1 Test set**

In order to evaluate the performance of the solution in each coordinate system (with respect to stability, precision and execution time) and to validate the method presented above, we used an extended test set of geodesics, which is available in GeographicLib (Karney 2016). This is a set of 500000 geodesics for the WGS84 ellipsoid of revolution, with *a* = 6378137 m and *f* = 1/298.257223563. The geodesics of the set are distributed into nine groups, as described in Table 1.

Table 1 Description of the geodesics in the test set

| Group | Identification number (ID) | Case |
|---|---|---|
| 1 | 1 – 100000 | randomly distributed |
| 2 | 100001 – 150000 | nearly antipodal |
| 3 | 150001 – 200000 | short distances |
| 4 | 200001 – 250000 | one end near a pole |
| 5 | 250001 – 300000 | both ends near opposite poles |
| 6 | 300001 – 350000 | nearly meridional |
| 7 | 350001 – 400000 | nearly equatorial |
| 8 | 400001 – 450000 | running between vertices ($\alpha_0 = \alpha_1 = 90°$) |
| 9 | 450001 – 500000 | ending close to vertices |

Each geodesic of the test set is defined by the data and the results for the direct geodesic problem, as described in Table 2.

Table 2 Description of data and results for the direct geodesic problem in the test set

| Quantity | Symbol | Unit | Accuracy |
|---|---|---|---|
| latitude at point 0 | $\varphi_0$ | degrees | exact |
| longitude at point 0 | $\lambda_0$ | degrees | exact, always 0 |
| azimuth at point 0 | $\alpha_0$ | clockwise from north in degrees | exact |
| latitude at point 1 | $\varphi_1^K$ | degrees | accurate to $10^{-18}$ deg |
| longitude at point 1 | $\lambda_1^K$ | degrees | accurate to $10^{-18}$ deg |
| azimuth at point 1 | $\alpha_1^K$ | degrees | accurate to $10^{-18}$ deg |
| geodesic distance from point 0 to point 1 | $s_{01}$ | meters | exact |

The values of $\varphi_1^K$, $\lambda_1^K$ and $\alpha_1^K$ were computed by Karney using high-precision direct geodesic calculations with the given $\varphi_0$, $\lambda_0$, $\alpha_0$ and $s_{01}$. For simplicity and without loss of generality, $\varphi_0$ is chosen in [0°, 90°], $\lambda_0$ is taken to be zero, $\alpha_0$ is chosen in [0°, 180°]. Furthermore, $\varphi_0$ and $\alpha_0$ are taken to be multiples of $10^{-12}$ deg and $s_{01}$ is a multiple of 0.1 μm in [0 m, 20003931.4586254 m]. Also, the values for $s_{01}$ for the geodesics running between vertices are truncated to a multiple of 0.1 pm and this is used to determine point 1. Finally, these conditions result to having $\lambda_1$ in [0°, 180°] and $\alpha_1$ in [0°, 180°] (Karney 2016).

For each geodesic in the test set, the systems of first-order differential equations (Eqs. (9) and Eqs. (26)) were integrated using the fourth-order Runge-Kutta numerical method (see Hildebrand 1974, Butcher 1987) using several different values of the number of steps *n*.

All algorithms were coded in FORTRAN95, were compiled by the open-source GNU FORTRAN compiler (at Level 2 optimization) and were executed on a personal computer running a 64-bit operating system. The main characteristics of the hardware were: Intel Core i5-2430M CPU (clocked at 2.4 GHz) and 6 GB of RAM.

For the computations, we used an 8-byte floating point arithmetic, which provides a precision of 18 decimal digits. However, for the conversion of the data of the full set (Table 2) from geodetic to Cartesian coordinates ($x_0$, $y_0 = 0$, $z_0$, $x_1^K$, $y_1^K$, $z_1^K$) using Eqs. (1), as well as for the computation of Clairaut's constant $C_1^K$ at the end point using Eq. (16), a 16-byte arithmetic was employed, which provides a precision of 32 decimal digits. The same high-precision arithmetic was also used for several tests (e.g. Table 16).

Detailed results of solving the direct geodesic problem in both coordinate systems are presented in the following sections.

### 4.2 Solving the geodesics in geodetic coordinates

The performance of the proposed method, using geodetic coordinates, was evaluated through a sub-set of 338640 geodesics, having the composition shown in Table 3.

Table 3 Composition of the sub-set

| Group | Total number of geodesics |
|---|---|
| 1 | 97997 |
| 2 | 49441 |
| 3 | 46699 |
| 4 | 0 |
| 5 | 0 |
| 6 | 0 |
| 7 | 50000 |
| 8 | 47294 |
| 9 | 47209 |

This sub-set was formed considering the constrains presented in Thomas and Featherstone (2005), in order to avoid the instabilities caused by the singularities in the geodetic coordinates. In particular, the geodesics of the sub-set satisfy the following criteria: $\varphi_0 < 85°$, $|\varphi_1| < 85°$, $1° < \alpha_0 < 179°$ and $1° < \alpha_1 < 179°$.

The direct geodesic problem in geodetic coordinates is solved using the input data $\varphi_0$, $\lambda_0$, $\alpha_0$ and $s_{01}$. The results ($\varphi_1$, $\lambda_1$) at the end point are converted to Cartesian coordinates ($x_1, y_1, z_1$) because this transformation is simple and numerically stable. In addition, this permits a direct comparison with the corresponding results using Cartesian coordinates.

At any point along the geodesic, the difference $\delta C = C - C_0$ is computed and its maximum value $\max|\delta C|$ is recorded. We also record the $\max(\max|\delta C|)$ of the whole set.

At the end point of each geodesic, we compute the difference $\delta r_1 = \left[(x_1 - x_1^K)^2 + (y_1 - y_1^K)^2 + (z_1 - z_1^K)^2\right]^{1/2}$ and we record the $\max(\delta r_1)$ of the whole set. Similarly, we compute the differences $\delta \alpha_1 = \alpha_1 - \alpha_1^K$ and $\delta C_1 = C_1 - C_1^K$ and we record the $\max|\delta \alpha_1|$ and the $\max|\delta C_1|$.

We note that we do not compute the azimuth at the intermediate points of the geodesic (Eq. (13)), since we have no other similar data to compare with.

All of the above data, along with the ID of the relevant geodesic, are presented in Table 4, for several values of the number of steps of the integration.

Table 4 Performance of the method on the subset of 338640 geodesics using geodetic coordinates

| n | $\max(\max\|\delta C\|)$ (m) | ID | $\max(\delta r_1)$ (m) | ID | $\max\|\delta \alpha_1\|$ (arcsec) | ID | $\max\|\delta C_1\|$ (m) | ID | t (s) |
|---|---|---|---|---|---|---|---|---|---|
| 1000 | 42 | 110733 | $1.5\,10^4$ | 145577 | 81 | 145577 | $3.2\,10^2$ | 110733 | 302 |
| 2000 | 2.7 | 110733 | $3.0\,10^2$ | 145577 | 1.7 | 145577 | $1.5\,10^1$ | 110733 | 618 |
| 5000 | $3.4\,10^{-2}$ | 110733 | 4.4 | 145577 | $2.3\,10^{-2}$ | 145577 | $3.7\,10^{-1}$ | 110733 | 1551 |
| 10000 | $1.1\,10^{-3}$ | 110733 | $1.7\,10^{-1}$ | 140863 | $7.4\,10^{-4}$ | 145577 | $2.2\,10^{-2}$ | 110733 | 3051 |
| 20000 | $3.5\,10^{-5}$ | 110733 | $9.1\,10^{-3}$ | 140863 | $2.2\,10^{-5}$ | 145577 | $1.4\,10^{-3}$ | 110733 | 5659 |
| 50000 | $3.6\,10^{-7}$ | 110733 | $2.2\,10^{-4}$ | 140863 | $2.2\,10^{-7}$ | 110733 | $3.5\,10^{-5}$ | 110733 | 14768 |
| 100000 | $2.8\,10^{-8}$ | 170970 | $1.4\,10^{-5}$ | 140863 | $1.3\,10^{-8}$ | 110733 | $2.2\,10^{-6}$ | 110733 | 28792 |
| 150000 | $4.7\,10^{-8}$ | 154108 | $2.7\,10^{-6}$ | 140863 | $5.6\,10^{-9}$ | 154108 | $4.3\,10^{-7}$ | 110733 | 43621 |

### 4.3 Solving the geodesics in Cartesian coordinates

In a similar manner, Table 5 presents the corresponding results of the performance of the proposed method in Cartesian coordinates, using the same sub-set of 338640 geodesics and using several values for the number of steps in the numerical integration. The input data are now $x_0$, $y_0$, $z_0$, $\alpha_0$ and $s_{01}$.

Table 5 Performance of the method on the subset of 338640 geodesics using Cartesian coordinates

| $n$ | $\max(\max|\delta C|)$ (m) | ID | $\max(\delta r_1)$ (m) | ID | $\max|\delta\alpha_1|$ (arcsec) | ID | $\max|\delta C_1|$ (m) | ID | $t$ (s) |
|---|---|---|---|---|---|---|---|---|---|
| 500 | $2.7\,10^{-6}$ | 490402 | $2.6\,10^{-4}$ | 126166 | $9.6\,10^{-5}$ | 424861 | $2.7\,10^{-6}$ | 490402 | 75 |
| 1000 | $8.3\,10^{-8}$ | 434510 | $1.6\,10^{-5}$ | 126166 | $6.0\,10^{-6}$ | 424861 | $8.3\,10^{-8}$ | 434510 | 123 |
| 2000 | $2.6\,10^{-9}$ | 462032 | $1.0\,10^{-6}$ | 126166 | $3.7\,10^{-7}$ | 424861 | $2.6\,10^{-9}$ | 462032 | 226 |
| 5000 | $9.1\,10^{-10}$ | 159630 | $2.6\,10^{-8}$ | 418118 | $9.6\,10^{-9}$ | 499415 | $9.1\,10^{-10}$ | 159630 | 478 |
| 10000 | $1.4\,10^{-9}$ | 159630 | $2.2\,10^{-9}$ | 159630 | $6.3\,10^{-10}$ | 443838 | $1.4\,10^{-9}$ | 159630 | 928 |

Comparing the results presented in Table 4 and Table 5, it is remarkable that, using Cartesian coordinates, we achieve similar or better levels of accuracy with a much smaller number of integration steps than using geodetic coordinates. In addition, execution time in Cartesian coordinates is reduced by a factor about 60, for similar accuracy levels.

Table 6 describes the results obtained solving the full set of 500000 geodesics in Cartesian coordinates. We note that, at any point along each geodesic, we compute the absolute value of the function $S$ (Eq. (17)) and we record the $\max|S|$. Thus, in addition to the data presented in the previous tables, we also give the value of $\max(\max|S|)$ for the whole set.

Table 6 Performance of the method on the full set of 500000 geodesics

| $n$ | max (max $|\delta C|$) (m) | ID | max (max $|S|$) | ID | max ($\delta r_1$) (m) | ID | max $|\delta \alpha_1|$ (arcsec) | ID | max $|\delta C_1|$ (m) | ID | $t$ (s) |
|---|---|---|---|---|---|---|---|---|---|---|---|
| 100 | $8.3\,10^{-3}$ | 413757 | $1.4\,10^{-9}$ | 130433 | $1.6\,10^{-1}$ | 117312 | $4.3\,10^{3}$ | 293277 | $8.3\,10^{-3}$ | 413757 | 83 |
| 200 | $2.6\,10^{-4}$ | 407322 | $4.3\,10^{-11}$ | 130433 | $1.0\,10^{-2}$ | 117312 | $2.7\,10^{2}$ | 293277 | $2.6\,10^{-4}$ | 407322 | 106 |
| 500 | $2.7\,10^{-6}$ | 490402 | $4.5\,10^{-13}$ | 261766 | $2.6\,10^{-4}$ | 117312 | 6.9 | 293277 | $2.7\,10^{-6}$ | 490402 | 132 |
| 1000 | $8.3\,10^{-8}$ | 434510 | $1.0\,10^{-14}$ | 252658 | $1.6\,10^{-5}$ | 289531 | $4.3\,10^{-1}$ | 293277 | $8.3\,10^{-8}$ | 434510 | 217 |
| 2000 | $2.6\,10^{-9}$ | 462032 | $<10^{-14}$ | 296969 | $1.0\,10^{-6}$ | 286391 | $2.7\,10^{-2}$ | 293277 | $2.6\,10^{-9}$ | 462032 | 358 |
| 5000 | $9.1\,10^{-10}$ | 159630 | $<10^{-14}$ | 486112 | $2.6\,10^{-8}$ | 299239 | $7.0\,10^{-4}$ | 292563 | $9.1\,10^{-10}$ | 159630 | 789 |
| 10000 | $1.4\,10^{-9}$ | 159630 | $<10^{-14}$ | 159630 | $2.2\,10^{-9}$ | 159630 | $4.6\,10^{-5}$ | 292563 | $1.4\,10^{-9}$ | 159630 | 1502 |
| 20000 | $3.9\,10^{-9}$ | 193167 | $<10^{-14}$ | 193167 | $4.2\,10^{-9}$ | 193167 | $7.7\,10^{-6}$ | 293277 | $3.9\,10^{-9}$ | 193167 | 2971 |

From Table 6 we conclude that our results are in agreement with the results of Karney's method, within a few nanometers in the end position and a few micro-arcseconds in the end azimuth.

With regard to the execution time, these impressive results are obtained at an average rate of 0.3 μs per integration step, which corresponds to about 6 ms for a geodesic using 20000 points. However, for most practical applications, a smaller number of steps is quite adequate.

In order to study in detail the differences in the results at the end points, we subsequently present, in Tables 7 to 15, the main results, separately for each group.

Table 7 Performance of the method for Group 1

| $n$ | max ($\delta r_1$) (m) | max $|\delta \alpha_1|$ (arcsec) | max $|\delta C_1|$ (m) | $t$ (s) |
|---|---|---|---|---|
| 100 | $1.6\,10^{-1}$ | $2.4\,10^{-1}$ | $7.7\,10^{-3}$ | 9 |
| 200 | $1.0\,10^{-2}$ | $1.5\,10^{-2}$ | $2.4\,10^{-4}$ | 12 |
| 500 | $2.6\,10^{-4}$ | $3.8\,10^{-4}$ | $2.5\,10^{-6}$ | 21 |
| 1000 | $1.6\,10^{-5}$ | $2.4\,10^{-5}$ | $7.7\,10^{-8}$ | 34 |
| 2000 | $1.0\,10^{-6}$ | $1.5\,10^{-6}$ | $2.4\,10^{-9}$ | 63 |
| 5000 | $2.6\,10^{-8}$ | $3.8\,10^{-8}$ | $5.1\,10^{-11}$ | 155 |
| 10000 | $1.7\,10^{-9}$ | $2.4\,10^{-9}$ | $6.3\,10^{-11}$ | 306 |
| 20000 | $3.2\,10^{-10}$ | $2.2\,10^{-10}$ | $9.5\,10^{-11}$ | 553 |

Table 8 Performance of the method for Group 2

| $n$ | max ($\delta r_1$) (m) | max $|\delta \alpha_1|$ (arcsec) | max $|\delta C_1|$ (m) | $t$ (s) |
|---|---|---|---|---|
| 100 | $1.6\,10^{-1}$ | $1.3\,10^{-1}$ | $8.3\,10^{-3}$ | 5 |
| 200 | $1.0\,10^{-2}$ | $8.2\,10^{-3}$ | $2.6\,10^{-4}$ | 6 |
| 500 | $2.6\,10^{-4}$ | $2.1\,10^{-4}$ | $2.7\,10^{-6}$ | 10 |
| 1000 | $1.6\,10^{-5}$ | $1.3\,10^{-5}$ | $8.3\,10^{-8}$ | 17 |
| 2000 | $1.0\,10^{-6}$ | $8.2\,10^{-7}$ | $2.6\,10^{-9}$ | 31 |
| 5000 | $2.6\,10^{-8}$ | $2.1\,10^{-8}$ | $6.9\,10^{-11}$ | 77 |
| 10000 | $1.8\,10^{-9}$ | $1.3\,10^{-9}$ | $6.0\,10^{-11}$ | 150 |
| 20000 | $3.9\,10^{-10}$ | $8.9\,10^{-11}$ | $8.5\,10^{-11}$ | 277 |

Table 9 Performance of the method for Group 3

| $n$ | max $(\delta r_1)$ (m) | max $|\delta\alpha_1|$ (arcsec) | max $|\delta C_1|$ (m) |
|---|---|---|---|
| 100 | $1.5\,10^{-11}$ | $3.1\,10^{-13}$ | $1.6\,10^{-11}$ |
| 200 | $4.1\,10^{-11}$ | $3.6\,10^{-13}$ | $2.8\,10^{-11}$ |
| 500 | $9.7\,10^{-11}$ | $1.2\,10^{-12}$ | $6.5\,10^{-11}$ |
| 1000 | $1.6\,10^{-10}$ | $1.2\,10^{-12}$ | $1.3\,10^{-10}$ |
| 2000 | $4.6\,10^{-10}$ | $3.9\,10^{-12}$ | $4.5\,10^{-10}$ |
| 5000 | $1.1\,10^{-9}$ | $5.7\,10^{-12}$ | $9.1\,10^{-10}$ |
| 10000 | $2.2\,10^{-9}$ | $5.5\,10^{-12}$ | $1.4\,10^{-9}$ |
| 20000 | $4.2\,10^{-9}$ | $5.3\,10^{-11}$ | $3.9\,10^{-9}$ |

Table 10 Performance of the method for Group 4

| $n$ | max $(\delta r_1)$ (m) | max $|\delta\alpha_1|$ (arcsec) | max $|\delta C_1|$ (m) |
|---|---|---|---|
| 100 | $1.6\,10^{-1}$ | $2.2\,10^{-3}$ | $1.2\,10^{-6}$ |
| 200 | $9.9\,10^{-3}$ | $1.4\,10^{-4}$ | $3.6\,10^{-8}$ |
| 500 | $2.5\,10^{-4}$ | $3.6\,10^{-6}$ | $3.7\,10^{-10}$ |
| 1000 | $1.6\,10^{-5}$ | $2.2\,10^{-7}$ | $1.2\,10^{-11}$ |
| 2000 | $9.9\,10^{-7}$ | $1.4\,10^{-8}$ | $1.0\,10^{-11}$ |
| 5000 | $2.5\,10^{-8}$ | $3.6\,10^{-10}$ | $1.5\,10^{-11}$ |
| 10000 | $1.6\,10^{-9}$ | $2.2\,10^{-11}$ | $2.3\,10^{-11}$ |
| 20000 | $2.6\,10^{-10}$ | $1.8\,10^{-11}$ | $2.9\,10^{-11}$ |

Table 11 Performance of the method for Group 5

| $n$ | max $(\delta r_1)$ (m) | max $|\delta\alpha_1|$ (arcsec) | max $|\delta C_1|$ (m) |
|---|---|---|---|
| 100 | $1.6\,10^{-1}$ | $4.3\,10^{3}$ | $1.5\,10^{-6}$ |
| 200 | $1.0\,10^{-2}$ | $2.7\,10^{2}$ | $4.6\,10^{-8}$ |
| 500 | $2.6\,10^{-4}$ | $6.9$ | $4.7\,10^{-10}$ |
| 1000 | $1.6\,10^{-5}$ | $4.3\,10^{-1}$ | $1.6\,10^{-11}$ |
| 2000 | $1.0\,10^{-6}$ | $2.7\,10^{-2}$ | $9.7\,10^{-12}$ |
| 5000 | $2.6\,10^{-8}$ | $7.0\,10^{-4}$ | $1.5\,10^{-11}$ |
| 10000 | $1.8\,10^{-9}$ | $4.6\,10^{-5}$ | $1.9\,10^{-11}$ |
| 20000 | $3.5\,10^{-10}$ | $7.7\,10^{-6}$ | $3.0\,10^{-11}$ |
| 50000 | $4.9\,10^{-10}$ | $3.6\,10^{-6}$ | $4.7\,10^{-11}$ |

Table 12 Performance of the method for Group 6

| $n$ | max $(\delta r_1)$ (m) | max $|\delta\alpha_1|$ (arcsec) | max $|\delta C_1|$ (m) |
|---|---|---|---|
| 100 | $1.6\,10^{-1}$ | $17$ | $6.9\,10^{-7}$ |
| 200 | $1.0\,10^{-2}$ | $1.1$ | $2.2\,10^{-8}$ |
| 500 | $2.6\,10^{-4}$ | $2.7\,10^{-2}$ | $2.2\,10^{-10}$ |
| 1000 | $1.6\,10^{-5}$ | $1.7\,10^{-3}$ | $6.9\,10^{-12}$ |
| 2000 | $1.0\,10^{-6}$ | $1.1\,10^{-4}$ | $2.2\,10^{-13}$ |
| 5000 | $2.6\,10^{-8}$ | $2.7\,10^{-6}$ | $<10^{-14}$ |
| 10000 | $1.7\,10^{-9}$ | $1.9\,10^{-7}$ | $<10^{-14}$ |
| 20000 | $3.6\,10^{-10}$ | $2.5\,10^{-9}$ | $<10^{-14}$ |

Table 13 Performance of the method for Group 7

| $n$ | max $(\delta r_1)$ (m) | max $|\delta\alpha_1|$ (arcsec) | max $|\delta C_1|$ (m) |
|---|---|---|---|
| 100 | $1.6\,10^{-1}$ | $4.5\,10^{-7}$ | $8.3\,10^{-3}$ |
| 200 | $1.0\,10^{-2}$ | $2.8\,10^{-8}$ | $2.6\,10^{-4}$ |
| 500 | $2.6\,10^{-4}$ | $7.2\,10^{-10}$ | $2.7\,10^{-6}$ |
| 1000 | $1.6\,10^{-5}$ | $4.5\,10^{-11}$ | $8.3\,10^{-8}$ |
| 2000 | $1.0\,10^{-6}$ | $2.8\,10^{-12}$ | $2.6\,10^{-9}$ |
| 5000 | $2.6\,10^{-8}$ | $1.1\,10^{-13}$ | $6.4\,10^{-11}$ |
| 10000 | $1.7\,10^{-9}$ | $4.0\,10^{-14}$ | $7.1\,10^{-11}$ |
| 20000 | $3.3\,10^{-10}$ | $4.0\,10^{-14}$ | $1.0\,10^{-10}$ |

Table 14 Performance of the method for Group 8

| $n$ | max $(\delta r_1)$ (m) | max $|\delta\alpha_1|$ (arcsec) | max $|\delta C_1|$ (m) |
|---|---|---|---|
| 100 | $1.6\,10^{-1}$ | $36$ | $8.3\,10^{-3}$ |
| 200 | $1.0\,10^{-2}$ | $2.3$ | $2.6\,10^{-4}$ |
| 500 | $2.6\,10^{-4}$ | $5.8\,10^{-2}$ | $2.7\,10^{-6}$ |
| 1000 | $1.6\,10^{-5}$ | $3.6\,10^{-3}$ | $8.3\,10^{-8}$ |
| 2000 | $1.0\,10^{-6}$ | $2.3\,10^{-4}$ | $2.6\,10^{-9}$ |
| 5000 | $2.6\,10^{-8}$ | $5.8\,10^{-6}$ | $7.4\,10^{-11}$ |
| 10000 | $1.8\,10^{-9}$ | $3.6\,10^{-7}$ | $7.3\,10^{-11}$ |
| 20000 | $3.6\,10^{-10}$ | $3.7\,10^{-8}$ | $9.4\,10^{-11}$ |
| 50000 | $5.1\,10^{-10}$ | $2.0\,10^{-8}$ | $1.7\,10^{-10}$ |

Table 15 Performance of the method for Group 9

| $n$ | max $(\delta r_1)$ (m) | max $|\delta\alpha_1|$ (arcsec) | max $|\delta C_1|$ (m) |
|---|---|---|---|
| 100 | $1.6\,10^{-1}$ | 65 | $8.3\,10^{-3}$ |
| 200 | $1.0\,10^{-2}$ | 4.1 | $2.6\,10^{-4}$ |
| 500 | $2.6\,10^{-4}$ | $1.0\,10^{-1}$ | $2.7\,10^{-6}$ |
| 1000 | $1.6\,10^{-5}$ | $6.5\,10^{-3}$ | $8.3\,10^{-8}$ |
| 2000 | $1.0\,10^{-6}$ | $4.1\,10^{-4}$ | $2.6\,10^{-9}$ |
| 5000 | $2.6\,10^{-8}$ | $1.0\,10^{-5}$ | $6.9\,10^{-11}$ |
| 10000 | $1.8\,10^{-9}$ | $6.5\,10^{-7}$ | $6.3\,10^{-11}$ |
| 20000 | $4.1\,10^{-10}$ | $6.7\,10^{-8}$ | $9.0\,10^{-11}$ |
| 50000 | $4.6\,10^{-10}$ | $2.4\,10^{-9}$ | $1.5\,10^{-10}$ |

We remark that the execution times for groups 2 to 9 (all have 50000 geodesics) were almost identical, so we present execution times only for group 1 (100000 geodesics) and group 2.

In Table 9 one may notice that a very small number of integration steps is sufficient to provide accurate results in the case of very short geodesics. In this case, an increase in the number of steps leads to worse results, which are attributed to the effects of round-off errors.

Since the results for groups 5, 8 and 9 indicate a lower accuracy of the value max$|\delta\alpha_1|$, we use a larger number of steps (50000) but the improvement is small, especially for group 5.

In order to examine further the cause of this behavior, we solved a particularly ill-behaved geodesic (ID 294750) using high-precision arithmetic (32 digits) and greater numbers of integration steps. The results, which are presented in Table 16, show a full agreement with those of Karney's method.

Table 16 Comparisons with Karney's data for the geodesic 294750 using 32-digit arithmetic

| $n$ | max $(\delta r_1)$ (m) | max $\|\delta\alpha_1\|$ (arcsec) | $t$ (s) |
| --- | --- | --- | --- |
| 50000 | $2.7\,10^{-12}$ | $4.1\,10^{-8}$ | 0.6 |
| 100000 | $2.1\,10^{-13}$ | $2.6\,10^{-9}$ | 1.2 |
| 200000 | $6.0\,10^{-14}$ | $1.6\,10^{-10}$ | 2.5 |
| 500000 | $5.0\,10^{-14}$ | $4.1\,10^{-12}$ | 6.2 |
| 1000000 | $<10^{-14}$ | $2.5\,10^{-13}$ | 12.2 |
| 2000000 | $<10^{-14}$ | $1.0\,10^{-14}$ | 25.7 |

## 5 Concluding remarks

A numerical solution of the geodesic initial value problem in geodetic and Cartesian coordinates on an oblate spheroid has been presented. The real power of the proposed method is that it is universal, i.e. can be used for arbitrary flattening and can be generalized in the case of a triaxial ellipsoid. Also, by setting $e = 0$ in the formulations, the geodesic initial value problem and its numerical solution in geodetic and Cartesian coordinates on a sphere is obtained as a degenerate case.

Comparing the results of solving the geodesic initial value problem in the two coordinate systems, we conclude that only the solution in Cartesian coordinates is complete, i.e. it works in the entire range of input data, it is stable, precise, accurate and fast, so it is recommended for use, especially when a tracing of the geodesic line is required. The precision of the method depends on the number of the significant digits of the computer system being used. However, current computer systems of everyday use are adequate to achieve excellent results in a very short time.

Furthermore, employing higher precision arithmetic (larger number of decimal digits), the results obtained, using the proposed method in Cartesian coordinates, are directly comparable with the results of Karney's method and they constitute an independent validation of Karney's geodesic dataset.

We also investigate ways to improve the performance of the proposed method, with regard to the precision attained in relation to the required execution time. We are experimenting with different orders of the numerical integration method, a more detailed examination of the dependence on the number of integration steps, as well as with using a variable step algorithm. We are also working on the generalization of this method and its application to a triaxial ellipsoid.

For the sake of a complete theory, knowledge of a precise analytical solution of the problem, as formulated above in Cartesian coordinates, is of interest. In addition, this would enable obtaining directly the results of the direct geodesic problem. Recall that the direct geodesic problem, and thus this proposed solution, can contribute to the solution of the inverse problem.

Finally, we plan to apply this concept, i.e. to solve the problem in space rather than on a surface, to other curves of geodetic importance, such as the normal section curve, the curve of alignment, the great elliptic arc and the loxodrome, as well as in other suitable geodetic problems.

**References**

Bessel FW (1826) Über die Berechnung der geographischen Längen und Breiten aus geodätischen Vermessungen. Astronomische Nachrichten 4:241-254. doi:10.1002/asna.18260041601

Bowring BR (1983) The geodesic inverse problem. Bulletin Géodésique 57:109-120. doi:10.1007/BF02520917

Butcher JC (1987) The numerical analysis of ordinary differential equations: Runge-Kutta and general linear methods. Wiley, New York

Deakin RE, Hunter MN (2010) Geometric geodesy, Part B. Lecture Notes, School of Mathematical & Geospatial Sciences, RMIT University, Melbourne, Australia

Felski A (2011) Computation of the azimuth of the great circle in Cartesian coordinates. Annual of Navigation 18:45-53

Fox C (1987) An introduction to the calculus of variations. Dover, New York

Hildebrand FB (1974) Introduction to numerical analysis, 2nd ed. Dover, New York

Holmstrom JS (1976) A new approach to the theory of geodesics on an ellipsoid. Navigation, Journal of The Institute of Navigation 23:237-244. doi:10.1002/j.2161-4296.1976.tb00746.x

Jank W, Kivioja LA (1980) Solution of the direct and inverse problems on reference ellipsoids by point-by-point integration using programmable pocket calculators. Surveying and Mapping 40:325-337

Karney CFF (2013) Algorithms for geodesics. Journal of Geodesy 87:43-55. doi:10.1007/s00190-012-0578-z

Karney CFF (2016) GeographicLib. http://geographiclib.sourceforge.net/html/. Accessed 01 November 2016

Kivioja LA (1971) Computation of geodetic direct and indirect problems by computers accumulating increments from geodetic line elements. Bulletin Géodésique 99:55-63. doi:10.1007/BF02521679

Mai E (2010) A fourth order solution for geodesics on ellipsoids of revolution. Journal of Applied Geodesy 4:145-155. doi:10.1515/jag.2010.014

Panou G (2013) The geodesic boundary value problem and its solution on a triaxial ellipsoid. Journal of Geodetic Science 3:240-249. doi:10.2478/jogs-2013-0028

Panou G, Delikaraoglou D, Korakitis R (2013) Solving the geodesics on the ellipsoid as a boundary value problem. Journal of Geodetic Science 3:40-47. doi:10.2478/jogs-2013-0007

Pittman ME (1986) Precision direct and inverse solutions of the geodesic. Surveying and Mapping 46:47-54

Rainsford HF (1955) Long geodesics on the ellipsoid. Bulletin Géodésique 37:12-22. doi:10.1007/BF02527187

Rapp RH (1993) Geometric geodesy, Part II. Department of Geodetic Science and Surveying, The Ohio State University, Columbus, Ohio, USA

Robbins AR (1962) Long lines on the spheroid. Survey Review 16:301-309. doi:10.1179/sre.1962.16.125.301

Rollins CM (2010) An integral for geodesic length. Survey Review 42:20-26. doi:10.1179/003962609X451663

Saito T (1970) The computation of long geodesics on the ellipsoid by non-series expanding procedure. Bulletin Géodésique 98:341-373. doi:10.1007/BF02522166

Saito T (1979) The computation of long geodesics on the ellipsoid through Gaussian quadrature. Bulletin Géodésique 53:165-177. doi:10.1007/BF02521087

Sjöberg LE (2012) Solutions to the ellipsoidal Clairaut constant and the inverse geodetic problem by numerical integration. Journal of Geodetic Science 2:162-171. doi:10.2478/v10156-011-0037-4

Sjöberg LE, Shirazian M (2012) Solving the direct and inverse geodetic problems on the ellipsoid by numerical integration. Journal of Surveying Engineering 138:9-16. doi:10.1061/(ASCE)SU.1943-5428.0000061

Sodano EM (1965) General non-iterative solution of the inverse and direct geodetic problems. Bulletin Géodésique 75:69-89. doi:10.1007/BF02530662

Struik DJ (1961) Lectures on classical differential geometry, 2nd ed. Dover, New York

Thomas CM, Featherstone WE (2005) Validation of Vincenty's formulas for the geodesic using a new fourth-order extension of Kivioja's formula. Journal of Surveying Engineering 131:20-26. doi:10.1061/(ASCE)0733-9453(2005)131:1(20)

Tseng WK (2014) An algorithm for the inverse solution of geodesic sailing without auxiliary sphere. Journal of Navigation 67:825-844. doi:10.1017/S0373463314000228

Vermeer M (2015) Mathematical geodesy. Lecture Notes, School of Engineering, Aalto University, Espoo, Finland

Vincenty T (1975) Direct and inverse solutions of geodesics on the ellipsoid with application of nested equations. Survey Review 23:88-93. doi:10.1179/sre.1975.23.176.88